\documentclass[%
 reprint,
 amsmath,amssymb,
 aps,
]{revtex4-2}

\usepackage{graphicx}
\usepackage{hyperref}
\usepackage{siunitx}
\usepackage{dcolumn}
\usepackage{bm}

\begin{document}

\preprint{APS}

\title{
Electron beam transverse phase space tomography using nanofabricated wire scanners with submicrometer resolution
}

\author{Benedikt~Hermann$^{1,3}$}
\email{Benedikt.Hermann@psi.ch}
\author{Vitaliy~A.~Guzenko$^1$}
\author{Orell~R.~H\"urzeler$^1$}
\author{Adrian~Kirchner$^2$}
\author{Gian~Luca~Orlandi$^1$}
\author{Eduard~Prat$^1$}
\author{Rasmus~Ischebeck$^1$}
\affiliation{\ \\ 
$^1$Paul Scherrer Institut, \\5232 Villigen PSI, Switzerland \\
$^2$Friedrich-Alexander-Universität Erlangen-Nürnberg, \\91054 Erlangen, Germany \\
$^3$Institute of Applied Physics, University of Bern, \\3012 Bern, Switzerland \\
}

\date{\today}

\begin{abstract}
Characterization and control of the transverse phase space of high-brightness electron beams is required at free-electron lasers or electron diffraction experiments for emittance measurement and beam optimization as well as at advanced acceleration experiments. Dielectric laser accelerators or plasma accelerators with external injection indeed require beam sizes at the micron level and below. We present a method using nano-fabricated metallic wires oriented at different angles to obtain projections of the transverse phase space by scanning the wires through the beam and detecting the amount of scattered particles. Performing this measurement at several locations along the waist allows assessing the transverse distribution at different phase advances. By applying a novel tomographic algorithm the transverse phase space density can be reconstructed. Measurements at the ACHIP chamber at SwissFEL confirm that the transverse phase space of micrometer-sized electron beams can be reliably characterized using this method. 
\end{abstract}

\maketitle
\section{\label{sec:intro}Introduction}
\begin{figure*}
\includegraphics[width=\textwidth]{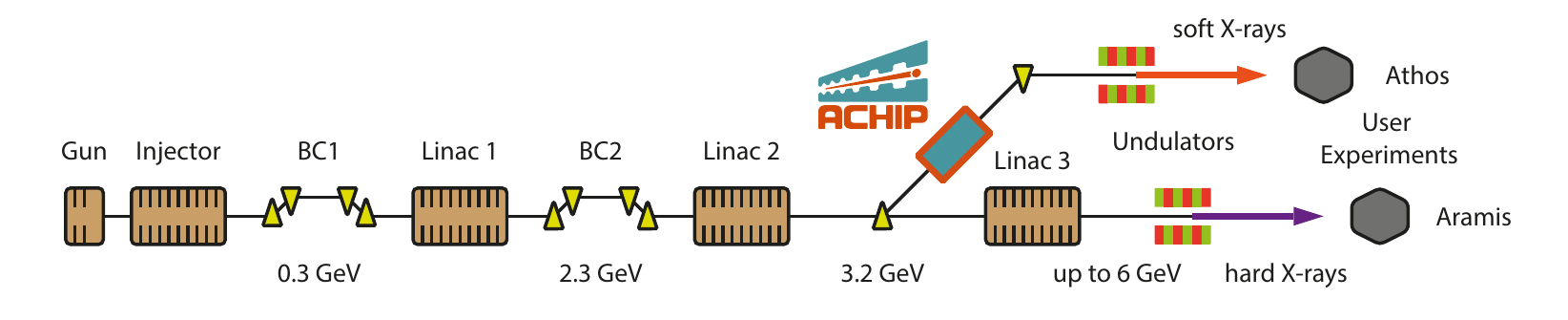}
\caption{\label{fig:sf} Schematic of the free-electron laser SwissFEL at PSI. The ACHIP chamber is located in the switch-yard to the Athos beamline at a beam energy of \SI{3.2}{GeV}.}
\end{figure*}
High-gradient advanced accelerator concepts including plasma and dielectric structure based schemes are developed at various laboratories for future compact accelerators.\\
The wavelength of the accelerating field in a plasma accelerator is given by the plasma wavelength which is typically on the order of tens of micrometers~\cite{esarey2009physics}. A Dielectric laser accelerator (DLA) is operating in the optical to near-infrared spectrum leading to structure apertures on the order of a single micrometer~\cite{england2014dielectric}.
Hence, suitable test beams for external injection have to be generated and characterized down to the sub-micrometer level. \\
Future compact free-electron laser facilities operating at small normalized emittances on the order of \SI{50}{\nm\radian}~\cite{rosenzweig2020} require profile monitors with micrometer resolution.
Electron diffraction requires an even smaller emittance to achieve the required coherence~\cite{ji2019ultrafast}.\\
Conventional beam profile monitors for ultra-relativistic electron beams are scintillating screens, optical transition radiation (OTR) screens and wire scanners. Screen-based methods provide single-shot two-dimensional information, whereas conventional wire scanners provide multi-shot one-dimensional information.\\
The thickness of the scintillating screen, the imaging lens and the camera pixel size limit the resolution of this method to around \SI{5}{\micro m} to \SI{10}{\micro m}~\cite{maxson2017direct, ischebeck2015transverse}. 
OTR screens with sub-micrometer resolution have been demonstrated, but their application is limited to uncompressed electron bunches ~\cite{bolzon2015very}.
Typical wire scanners at free-electron laser facilities consist of cylindrical metallic (aluminum or tungsten) wires with diameters down to \SI{5}{\micro\m}~\cite{orlandi2016design}. Projections of the transverse beam distribution can be measured by moving the stretched wire through the beam and correlating the wire position to the signal of a downstream beam loss monitor, which detects the scattered particle shower. 
Recent developments at PSI and FERMI led to single (one-dimensional) wire scanners fabricated with electron beam lithography reaching sub-micrometer resolution~\cite{orlandi2020nanofabricated, borrelli2018generation}.
Based on this technology we designed a wire scanner consisting of nine wires arranged radially at different angles, as a tool for precise beam profile tomography at the ACHIP (Accelerator on a Chip International Program) interaction chamber, which is installed in the Athos branch of SwissFEL at PSI (see Fig.~\ref{fig:sf}). This chamber is planned to support DLA research and development~\cite{prat2017outline, ferrari2018achip}. A possible application of DLA technology for FELs is the generation of a micro-bunched pulse train using laser-based energy modulation followed by magnetic compression~\cite{hermann2019laser}. \\  
The electrons at the ACHIP interaction point at SwissFEL possess a mean energy of \SI{3.2}{GeV} and are strongly focused by an in-vacuum permanent magnet triplet~\cite{prat2017outline}. 
A six-dimensional positioning system (hexapod) at the center of the chamber is used to exchange, align, and scan samples or a wire scanner for diagnostics.\\
In this manuscript, we demonstrate that the transverse phase space of a focused electron beam can be precisely characterized with a series of wire scans at different angles and locations along the waist. The transverse phase space ($x-x'$ and $y-y'$) is reconstructed with a novel particle-based tomographic algorithm. This technique goes beyond conventional one-dimensional wire scanners since it allows us to asses the four-dimensional transverse phase space.
We apply this algorithm to a set of wire scanner measurements performed with nano-fabricated wires at the ACHIP chamber at SwissFEL and reconstruct the dynamics of the transverse phase space of the focused electron beam along the waist.

\section{\label{sec:setup}Experimental Setup}

\subsection{Accelerator Setup}
The generation and characterization of a micro-meter sized electron beam in the ACHIP chamber at SwissFEL requires a low-emittance electron beam. 
The beam size along the accelerator is given by:
\begin{equation}
    \sigma(z) = \sqrt{\beta(z) \varepsilon_n(z)/ \gamma(z)},
\end{equation}
where $\beta$ denotes the Twiss (or Courant-Snyder) parameter of the magnetic lattice, $\gamma$ is the relativistic Lorentz factor of the electrons and $\varepsilon_n$ is the normalized emittance of the beam. 
With an optimized lattice a minimal $\beta$-function of around \SI{1}{cm} in the horizontal and \SI{1.8}{cm} in the vertical plane is expected from simulations~\cite{prat2017outline, ferrari2018achip}.\\
In order to reduce chromatic effects of the focusing quadrupoles~\cite{Mostacci2012}, we minimize the projected energy spread by accelerating the beam in most parts of the machine close to on-crest acceleration. From simulations, we expect an optimized projected energy spread of \SI{42}{keV} for a \SI{3}{GeV}-beam with a charge of \SI{1}{pC}~\cite{prat2017outline}, which corresponds to a relative energy spread of $\SI{1.4e-5}{}$. For this uncompressed and low-energy-spread beam we expect chromatic enlargement of the focused beam size on the order of \SI{0.1}{\percent}. \\
To lower the emittance of the beam, the bunch charge is reduced to approximately \SI{1}{pC} from the nominal bunch charge at SwissFEL (\SIrange{10}{200}{pC}). The laser aperture and pulse energy at the photo-cathode, as well as the current of the gun solenoid, are empirically tuned to minimize the emittance for the reduced charge. 
The emittance is measured at different locations along the accelerator with a conventional quadrupole scan~\cite{prat2014symmetric} and a scintillating YAG:Ce screen. After the second bunch compressor, which is the last location for emittance measurements before the ACHIP chamber, the normalized horizontal and vertical emittances are found to be \SI{93}{\nm\radian} and \SI{157}{\nm\radian} with estimated uncertainties below \SI{10}{\percent}. The difference between the horizontal and vertical emittance could be the result of an asymmetric laser spot on the cathode. The electron energy at this emittance measurement location is \SI{2.3}{GeV}.
Subsequently, the beam is accelerated further to \SI{3.2}{GeV} and directed to the Athos branch by two resonant deflecting magnets (kickers) and a series of dipole magnets~\cite{abela2019swissfel}. Finally, the beam is transported to the beam stopper upstream of the Athos undulators.

\subsection{ACHIP Chamber}
\begin{figure}
\includegraphics[width=0.48\textwidth]{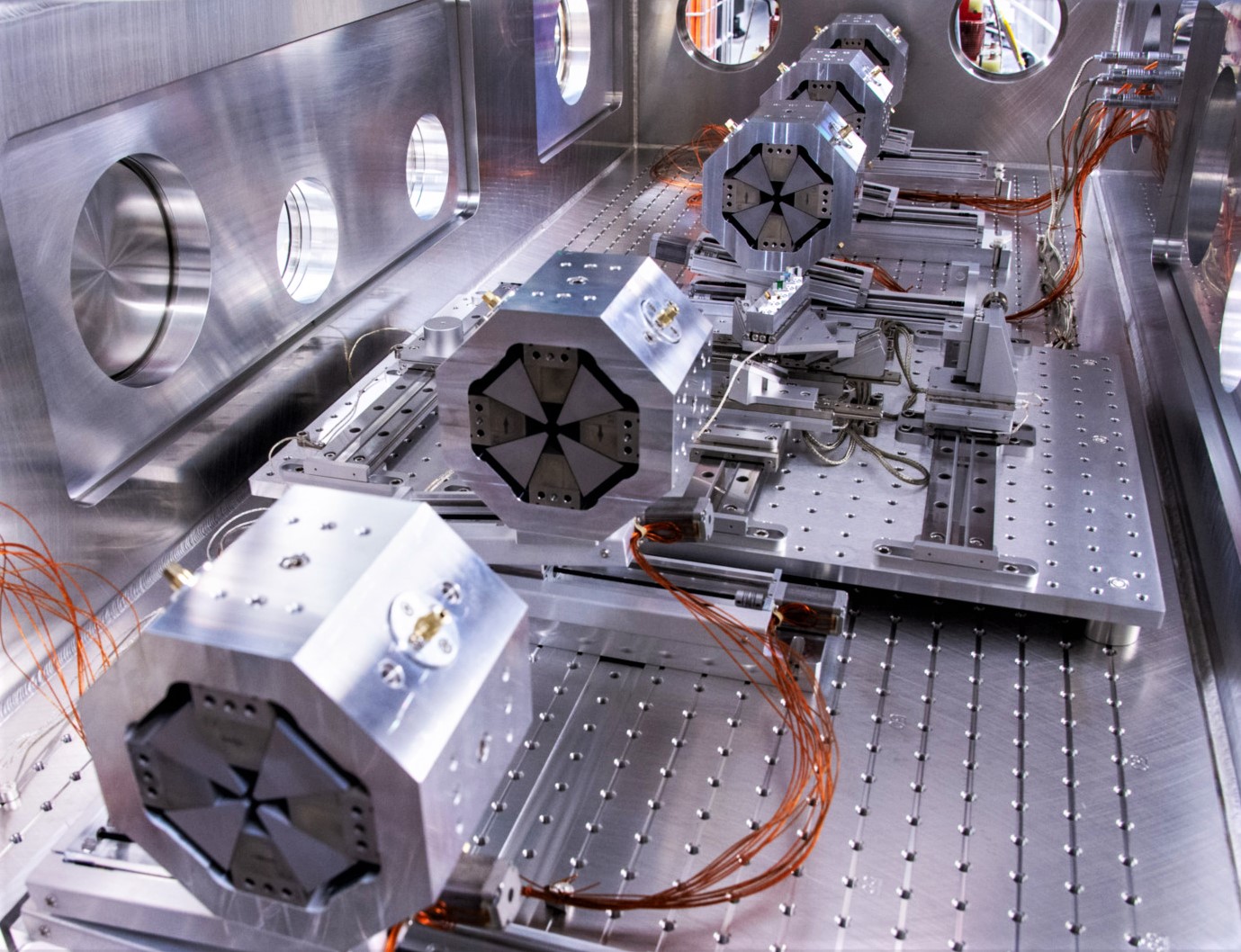}
\caption{\label{fig:chamber} Inside view of the ACHIP chamber. Movable quadrupoles for focusing and re-matching are seen in the front and back. The hexapod for sample positioning is located at the center. Image adapted from~\cite{Ischebeck:2020yvw} under \href{https://creativecommons.org/licenses/by/3.0/}{Creative Commons Attribution 3.0 licence}.}
\end{figure}

The ACHIP chamber at SwissFEL is a multi-purpose test chamber, designed and built for DLA research. It is located in the switch-yard of SwissFEL, where the electron beam has an energy of around \SI{3.2}{GeV}. The electron beam is focused by an in-vacuum quadrupole triplet and matched back by a second symmetric quadrupole triplet. All six magnets can be remotely retracted from the beam line for standard SwissFEL operation. The positioning system allows the alignment of the quadrupoles with respect to the electron beam. The magnetic center of the quadrupole is found by observing and reducing transverse kicks with a downstream screen or beam position monitor.
At the center of the chamber a hexapod allows positioning different samples in the electron beam path. Figure~\ref{fig:chamber} shows the interior of the ACHIP chamber including the permanent magnets and the hexapod. Further details about the design of the experimental chamber can be found in~\cite{prat2017outline, ferrari2018achip} and the first results of the beam characterization can be found in~\cite{Ischebeck:2020yvw}.

\subsection{\label{sec:ws}Nano-Fabricated Wire Scanner}
\begin{figure}
\includegraphics[width=0.48\textwidth]{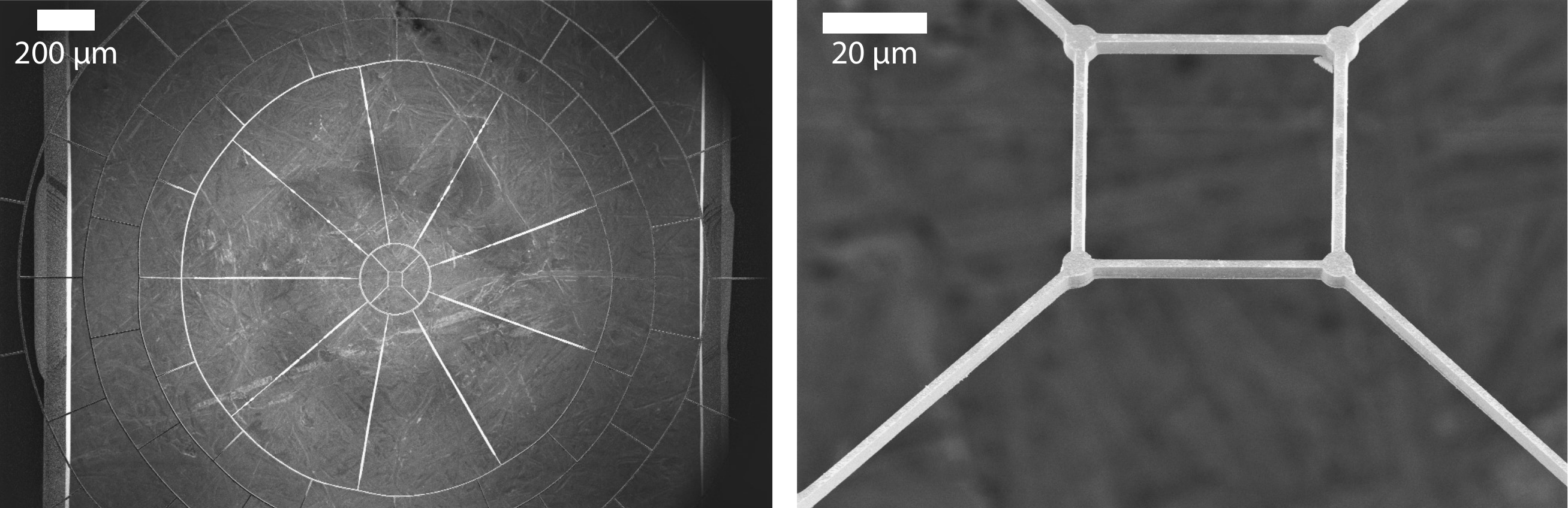}
\caption{\label{fig:ws} Scanning electron microscope images of the free-standing wire scanner device. Nine radial wires which are used for the wire scans are supported by a spiderweb-shaped structure attached to a silicon frame. At the center of the geometry a square simplifies the alignment of the wire scanner with respect to the electron beam. Scanning the square horizontally and vertically across the beam provides 4 distinct peaks (for beam sizes smaller than \SI{50}{\micro \meter}). The center of the geometry can be referenced to the hexapod coordinate system from the location of these peaks.}
\end{figure}

Nano-fabricated wires are installed on the hexapod for the characterization of the focused beam profile. The wire scan device consists of nine free-standing \SI{1}{\micro m} wide metallic (Au) stripes. The nine radial wires are supported by a spiderweb-shaped structure attached to a silicon frame. 
A scanning electron microscope image of the wire scanner sample is shown in Fig.~\ref{fig:ws}. 
We chose nine homogeneously spaced wires for our design, since this configuration allows us to access any wire angle within the tilt limits of the hexapod.
The sample was fabricated at the Laboratory for Micro and Nanotechnology at PSI by means of electron beam lithography. The \SI{1}{\micro m} wide stripes of gold are electroplated on a \SI{250}{nm} thick Si$_3$N$_4$ membrane, which is removed afterwards with a KOH bath. The fabrication process and performance for this type of wire scanner are described in detail in~\cite{orlandi2020nanofabricated}.
The hexapod moves the wire scan device on a polygon path to scan each of the nine wires orthogonally through the electron beam. Hereby, projections along different angles ($\theta$) of the transverse electron density can be measured. 
The two-dimensional transverse beam profile can be obtained using tomographic reconstruction techniques. The hexapod can position the wire scanner within a range of \SI{20}{cm} along the beam direction ($z$). By repeating the wire scan measurement at different locations around the waist, the transverse phase space and emittance of the beam can be inferred.

\subsection{\label{sec:blm}Beam Loss Monitor}
Electrons scatter off the atomic nuclei of the metallic wire and a particle shower containing mainly X-rays, electrons and positrons is generated. The intensity of the secondary particle shower depends on the electron density integrated along the wire and is measured with a downstream beam loss monitor (BLM). The BLM consists of a scintillating fiber wrapped around the beam pipe. The fiber is connected to a photo-multiplier tube (PMT). The signal of the PMT is read-out beam synchronously in a shot-by-shot manner. To avoid saturation of the PMT, the gain voltage needs to be set appropriately.
SwissFEL is equipped with a series of BLMs, which are normally used to detect unwanted beam losses and are connected to an interlock system. For the purpose of wire scan measurements, individual BLMs can be excluded from the machine protection system. Details about the BLMs at SwissFEL can be found in~\cite{cigdem2020lossmonitors}.
For the wire scan measurement reported here, a BLM located \SI{10}{m} downstream of the interaction with the wire was used. 

\section{\label{sec:alg}Transverse Phase Space Reconstruction Algorithm}
Inferring a density distribution from a series of projection measurements is a problem arising in many scientific and medical imaging applications. 
Standard tomographic reconstruction techniques, e.g., filtered back projection or algebraic reconstruction technique~\cite{gordon1970algebraic} use an intensity on a grid to represent the density to be reconstructed. The complexity of these algorithms scales as $\mathcal{O}(n^d)$, where $n$ is the number of pixels per dimension and $d$ is the number of dimensions of the reconstructed density. Typically, for real space density reconstruction, $d$ is 2 (slice reconstruction) or 3 (volume reconstruction). 
In the case of transverse phase space tomography $d$ equals 4 ($x, x', y, y'$), leading to very long reconstruction times. \\
We developed a reconstruction algorithm based on a macro-particle distribution (instead of the intensity on grid), where each macro-particle, from now on called particle, represents a point in the four-dimensional phase space. 
The complexity of this algorithm is proportional to $n_{p}$ (number of particles) and is independent on the dimension of the reconstruction domain. 
The particle density is then given by applying a Gaussian kernel to each coordinate of the particle ensemble:
\begin{equation}
    G_{\kappa} = \frac{1}{\sqrt{2\pi}\rho_\kappa} \exp \left( - \frac{\kappa^2}{2 \rho_\kappa^2} \right), \kappa \in \{x,x',y,y'\}
    \label{eq:kernel}
\end{equation}
where we choose $\rho_{x', y'} = \rho_{x,y} / z_{\mathrm{max}}$, with $z_{\mathrm{max}}$ the range of the measurement along $z$. 
Choosing the right kernel size is important for an appropriate reconstruction of the beam. It is dimensioned such that $\rho_{x,x',y,y'}$ represents the length scale below which we expect only random fluctuations in the particle distribution, which are not reproducible from shot to shot. Note that despite the Gaussian kernel, this reconstruction does not assume a Gaussian distribution of the beam, but is able to reconstruct arbitrary distributions that vary on a length scale given by $\rho_{x,x',y,y'}$.\\
The ensemble of particles is iteratively optimized so that their projections match with the set of measured projections. 
The algorithm starts from a homogeneous particle distribution. 
One iteration consists of the following operations. 
\begin{itemize}
  \item Transport $T(z)$
  \item Rotation $R(\theta)$
  \item Histogram of the transported and rotated coordinates
  \item Convolution with wire profile
  \item Interpolation to measured wire positions
  \item Comparison of reconstruction and measurement
  \item Redistribution of particles
\end{itemize}
In the case of ultra-relativistic electrons transverse space charge effects can be neglected since they scale as $\mathcal{O}(\gamma^{-2})$ and hence $T(z)$ becomes the ballistic transport matrix: 
\begin{equation}
T(z) = \begin{pmatrix}
1 & z \\
0 & 1 
\end{pmatrix}	
\end{equation}
for ($x, x'$) and ($y, y'$).
The rotation matrix is then applied to ($x, y$):
\begin{equation}
R(\theta) = \begin{pmatrix}
\cos \theta & \sin \theta \\
-\sin \theta & \cos \theta 
\end{pmatrix}.
\end{equation}
Afterwards, the histogram of the particles' transported and rotated $x$ coordinates is calculated. Note that the bin width needs to be smaller than the width of the wire, to ensure an accurate convolution with the wire profile. This becomes important when the beam size or beam features are smaller than the wire width.
Next, the convolution of the histogram and the wire profile is interpolated linearly to the measured wire positions $\xi$. Now, the reconstruction can be directly compared to the measurement: 
\begin{equation}
    \Delta_{z, \theta}(\xi) = \frac{P^{m}_{z, \theta}(\xi) - P^{r}_{z, \theta}(\xi)}{\max_\xi P^{r}_{z, \theta}(\xi)}, 
\end{equation}
where $P^{m}_{z, \theta}$ and $P^{r}_{z, \theta}$ are the measured and reconstructed projections for the current iteration at position $z$ and  angle $\theta$.
The difference between both profiles quantifies over- and under-dense regions in the projection.
Then, $\Delta_{z, \theta}(\xi)$ is interpolated back to the particle coordinates along the wire scan direction, yielding $\Delta_{z, \theta}^i$ for the $i$-th particle. 
Afterwards, we calculate the average over all measured $z$ and $\theta$:
\begin{equation}
\Delta^i = \frac{1}{n_\theta n_z} \sum_{\theta, z} \Delta_{z, \theta}^i.
\end{equation}
The sign of $\Delta^i$ indicates if a particle is located in an over- or under-dense region represented by the current particle distribution. According to the magnitude of $\Delta^i$ the new particle ensemble is generated. A particle is copied or removed from the previous distribution with a probability based on $|\Delta^i|$. This process is implemented by drawing a pseudo-random number $\chi^i \in [0,1[$ for each particle. In case $\chi^i < |\Delta^i|/s_{\mathrm{max}}$, particle $i$ is copied or removed from the distribution (depending on the sign of $\Delta^i$). Otherwise, the particle remains in the ensemble. Here, $s_{\mathrm{max}}$ is the maximum of all measured BLM signals and is used to normalize $\Delta^i$ for the comparison with $\chi^i \in [0,1[$. 
This process makes sure that particles in highly under-dense (over-dense) regions are created (removed) with an increased probability.\\
In the last step of each iteration, a small random value is added to each coordinate according to the Gaussian kernel defined in Eq.~\ref{eq:kernel}. This smoothens the distribution on the scale of $\rho$.
For the reconstruction of the measurement presented in Sec.~\ref{sec:res}, $\rho_{x,y}$ was set to \SI{80}{nm}.\\
The iterative algorithm is terminated by a criterion based on the relative change of the average of the difference $\Delta_{z, \theta}^i$ (further details in Appendix~\ref{app:term}).
The measurement range along $z$ ideally covers the waist and the spacing between measurements is reduced close to the waist, since the phase advance is the largest here.
Since the algorithm does not assume a specific shape (e.g., Gaussian) of the distribution, asymmetries, double-peaks, or halos of the distribution can be reconstructed (an example is shown in Appendix~\ref{app:nonGaussian}). 
Properties of the transverse phase space including, transverse emittance in both planes, astigmatism and Twiss parameters can be calculated from the reconstructed distribution.
To obtain the full 4D emittance, cross-plane information, such as correlations in $x-y'$ or $x'-y$ need to be assessed. For this purpose, the phase advance has to be scanned independently in both planes. This can be achieved with a multiple quadrupole scan as explained for instance in~\cite{Raimondi:1993fc, prat2014four} but is not achieved by measuring beam projections along a waist, as the phase advance in both planes is correlated.\\
The presented phase space reconstruction algorithm could also be adapted to use two-dimensional profile measurements from a screen at different phase advances to characterize the four-dimensional transverse phase space. \\
The python-code related to the described tomographic reconstruction technique is made available on github~\cite{GitHermannTomo}.

\subsection{Reconstruction of a Simulated Measurement}
\begin{figure}
\includegraphics[width=0.48\textwidth]{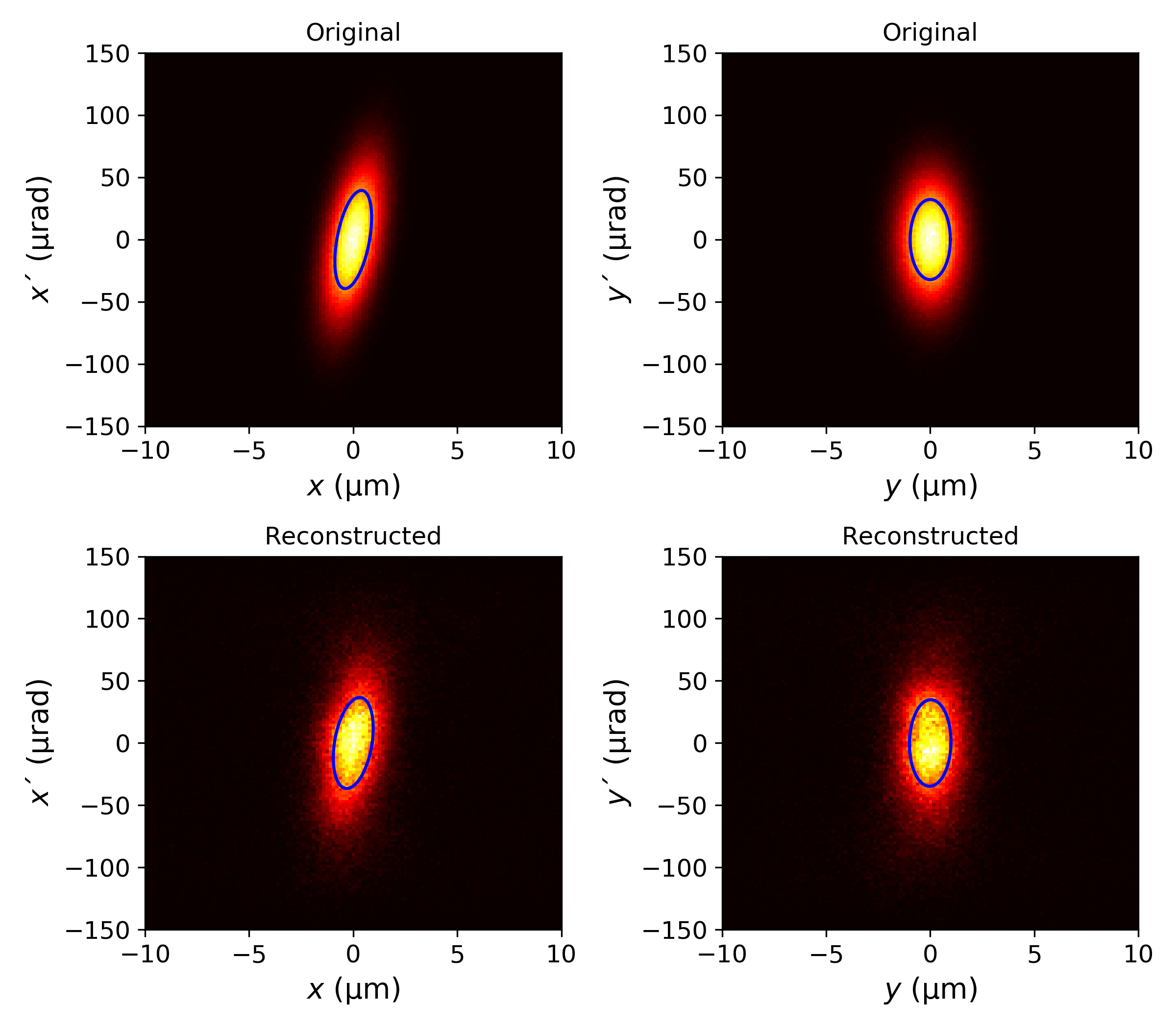}
\caption{\label{fig:simBeam} Reconstruction from simulated measurement. The original distribution in the transverse phase space is shown in the upper row. An astigmatism of \SI{-1}{cm} is added to the horizontal plane (tilt in $x-x'$). The algorithm reconstructs the transverse phase space based on a set of simulated wire scan projections. The result of the reconstruction is shown in the lower row. The $1\sigma$-ellipse of a 2D Gaussian fit is drawn in blue for each histogram.}
\end{figure}
To verify the reconstruction algorithm, we generate a test distribution and calculate a set of wire scan projections (nine projections along different angles at seven locations along the waist). The algorithm then reconstructs the distribution based on these simulated projections. For this test, we choose a Gaussian beam distribution with Twiss parameters $\beta_x^* = \SI{2.0}{cm}$, $\beta_y^* = \SI{3.0}{cm}$ and a transverse emittance of \SI{200}{\nm\radian} in both planes. An astigmatism of \SI{-1}{cm} (longitudinal displacement of the horizontal waist) is artificially introduced. Moreover, noise is added to the simulated wire scan profiles to obtain a signal-to-noise ratio similar to the experimental data show in Sec.~\ref{sec:res}. The Gaussian kernel size for the reconstruction $\rho_{x,y}$ (see Eq.~\ref{eq:kernel}) is \SI{80}{nm}, which is around one order of magnitude smaller than the beam size in this test.  Figure~\ref{fig:simBeam} compares the original and reconstructed transverse phase space at $z=\SI{0}{cm}$. Good agreement ($<\SI{10}{\percent}$ error) is achieved for the emittances and astigmatism, which is manifested as a tilt in the $x-x'$ plane.
For this numerical experiment, the algorithm terminates according to the criterion described in Appendix~\ref{app:term} after around 100 iterations. The run-time on a single-core of a standard personal computer is around two minutes. Parallelising the computation on several cores would reduce the computation time by few orders of magnitude. 

\section{\label{sec:res}Results}
\begin{figure*}
\includegraphics[width=\textwidth]{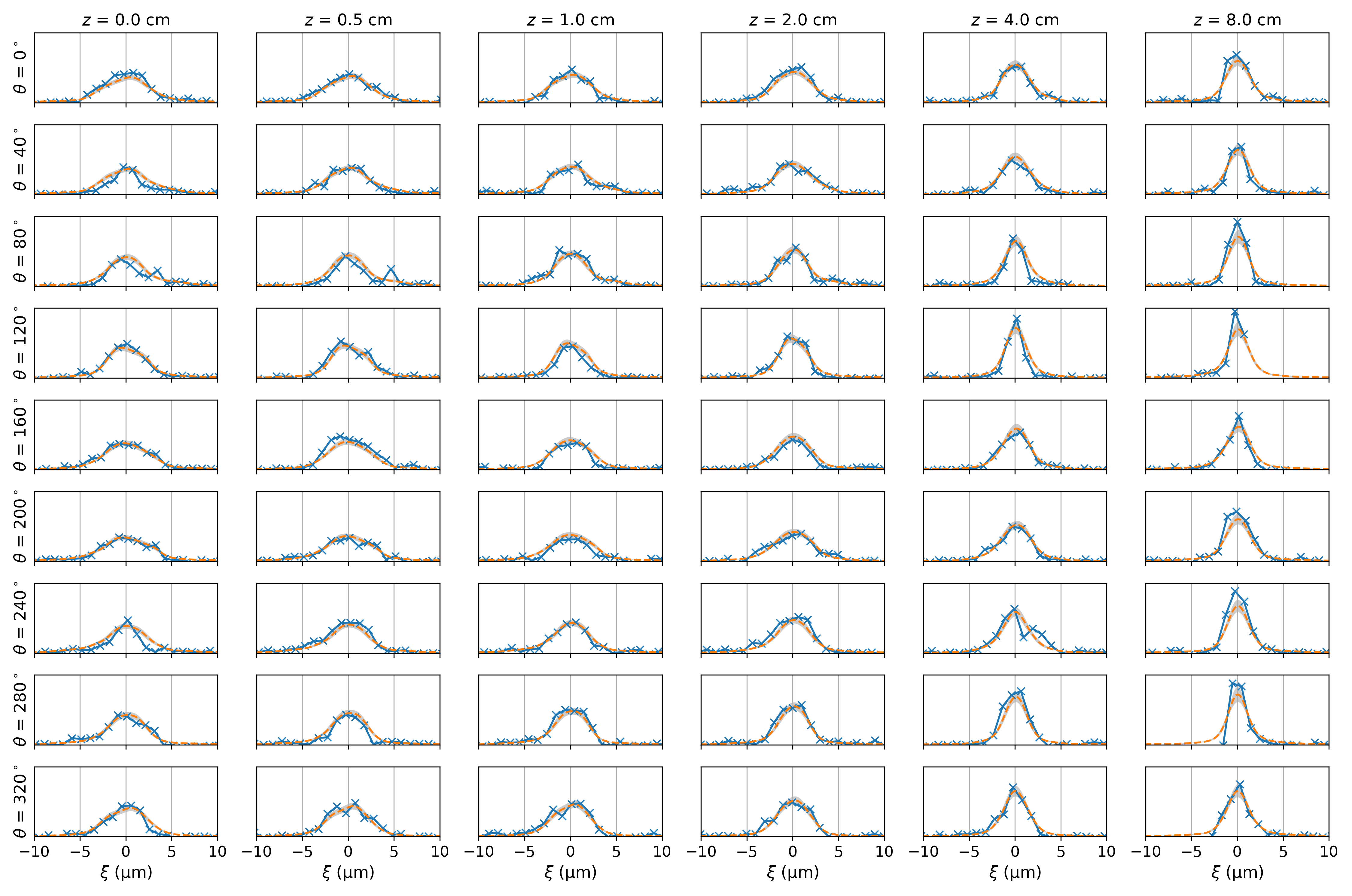}
\caption{\label{fig:projections} Measured (blue crosses) and reconstructed (orange dashed) profiles of the electron beam distribution. The vertical axes are identical for all sub-plots and show the BLM signal or reconstruction in arbitrary units. Sub-plots in the same column correspond to the same $z$ location of the wire scanner and sub-plots in the same row correspond to the same projection angle $\theta$. The grey area depicts the uncertainty of the reconstruction. For the last column ($z = \SI{8}{cm}$) the scan range did not cover the entire beam profile for all scans due to a misalignment of the electron propagation direction and the $z$-axis of the hexapod, which results in a transverse offset of the wire scanner device with respect to the electron beam. This effect is the largest for the last scan ($z = \SI{8}{cm}$) since the wire scanner was aligned to the beam axis at $z = \SI{0}{cm}$.}
\end{figure*}

\begin{figure*}
\includegraphics[width=\textwidth]{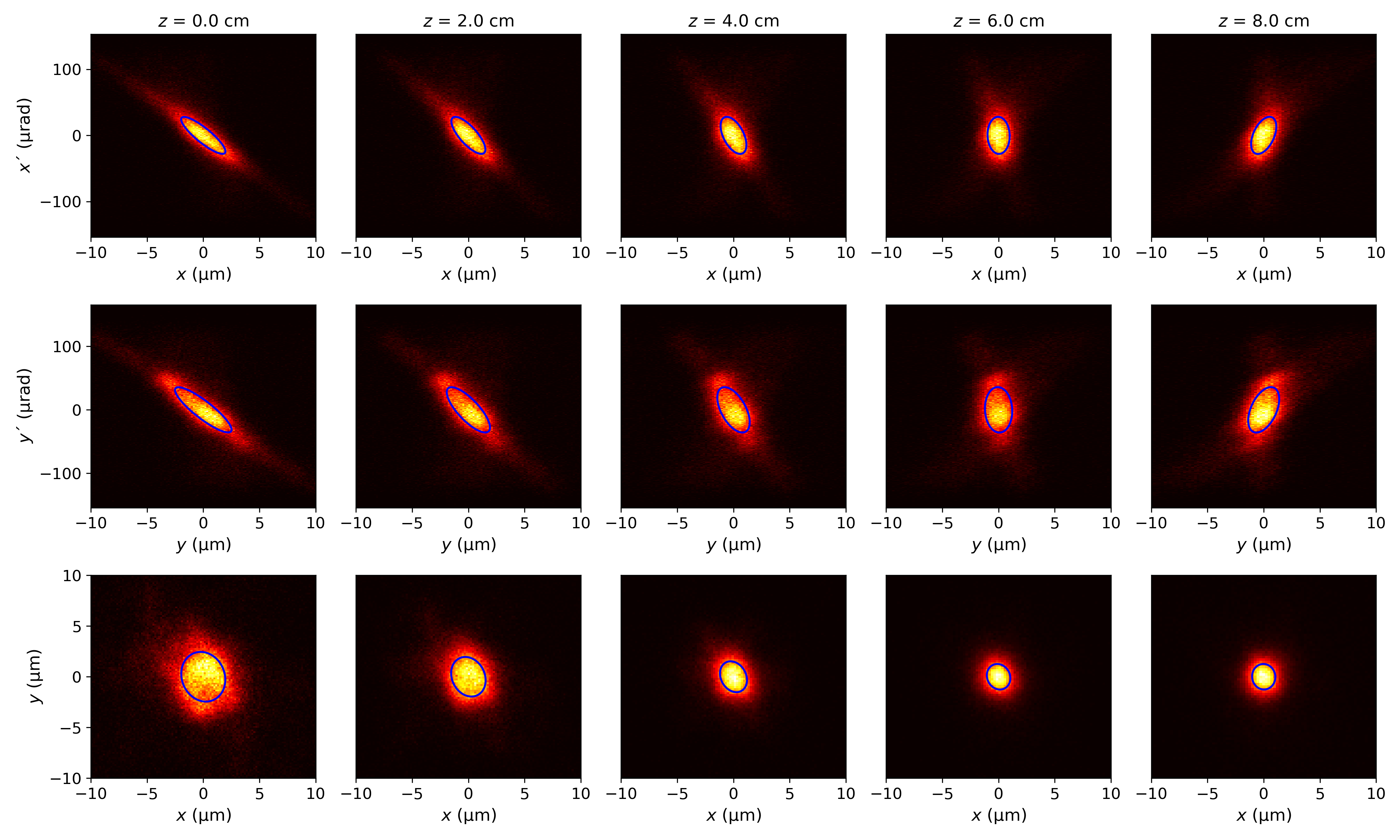}
\caption{\label{fig:phaseSpace} 2D histograms of the phase space reconstructed from wire scan measurements. Sub-plots in the same column correspond to the same $z$ location. The first row shows the $x-x'$ and the second row shows the $y-y'$ phase space. The last row depicts the corresponding beam profile ($x-y$). The $1\sigma$-ellipse of a 2D Gaussian fit is drawn in blue for each histogram.}
\end{figure*}

\begin{figure}
\includegraphics[width=0.45\textwidth]{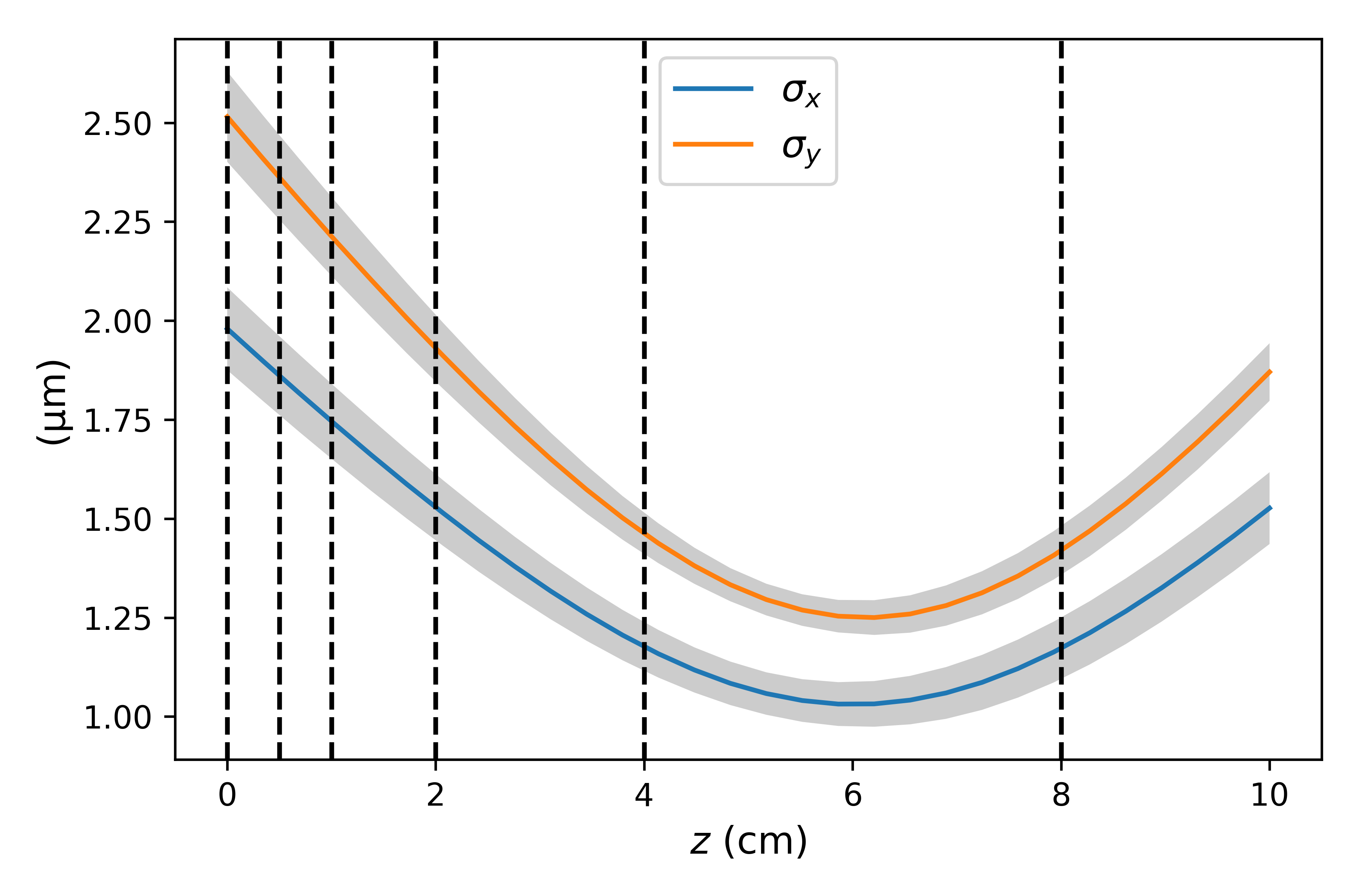}
\caption{\label{fig:sigmaEvolution} Evolution of the reconstructed beam size around the waist. The dashed vertical lines indicate wire scan measurement locations along $z$. At each of the six locations wire scans are carried out along nine different angles. Since we expected the waist to be around $z=\SI{0}{cm}$, the distance between measurements is reduced here.}
\end{figure}

We have measured projections of the transverse electron beam profile at the ACHIP chamber at SwissFEL with the accelerator setup,  wire scanner and BLM detector described in Sec.~\ref{sec:setup}. 
All nine wire orientations are used at six different locations along the waist of the electron beam. This results in a total of 54 projections of the electron beam's transverse phase space. Lowering the number of projections limits the possibility to observe inhomogeneities of the charge distribution. 
The distance between measurement locations is increased along $z$, since the expected waist location was around $z=\SI{0}{cm}$. All 54 individual profiles are shown in Fig.~\ref{fig:projections}. 
In each sub-plot, the orange dashed curve represents the projection of the reconstructed phase space for the respective angle $\theta$ and longitudinal position $z$. The reconstruction represents the average distribution over many shots and agrees with most of the measured data points.
Discrepancies arise due to shot-to-shot position jitter, charge fluctuations, or density variations of the electron beam. The effect of these error sources is discussed further in Appendix~\ref{app:err}.
The evolution of the reconstructed transverse phase space along the waist is depicted in Fig.~\ref{fig:phaseSpace}. The expected rotation of the transverse phase space around the waist is clearly observed. The position of the waist is found to be at around $z = \SI{6.2}{cm}$ downstream of the center of the chamber. Figure~\ref{fig:sigmaEvolution} shows the beam size evolution around the waist. 
We quantify the normalized emittance and $\beta$-function of the distribution by fitting a 2D Gauss function to the distribution in the ($x$, $x'$) and ($y$, $y'$) phase space. The 1-$\sigma$ ellipse of the fit is drawn in blue in all sub-plots of Fig.~\ref{fig:phaseSpace}. We use the following definition for the normalized emittance:
\begin{equation}
    \varepsilon_n = \gamma A_{1\sigma} / \pi,
\end{equation}
where $A_{1\sigma}$ is the area of the 1-$\sigma$ ellipse in transverse phase space. 
The values for the reconstructed emittance, minimal $\beta$-function ($\beta^*$) and beam size at the waist are summarized in Table~\ref{tab:emi}. The measurement range (\SI{8}{cm}) along the waist with $\beta^*=\SI{3.7}{cm}$ covers a phase advance of around \SI{90}{\degree}.
\begin{table}
    \centering
    \begin{tabular}{c|c|c|c}
            & $\varepsilon_n$ (\SI{}{\nm\radian})   & $\beta^*$ (\SI{}{cm}) & $\sigma^*$ (\SI{}{\micro m})\\ \hline
       x    & \SI{186(15)}{}                        & \SI{3.7(2)}{}     & \SI{1.04(6)}{} \\
       y    & \SI{278(18)}{}                        & \SI{3.7(2)}{}     & \SI{1.26(5)}{} 
    \end{tabular}
    \caption{Normalized emittance $\varepsilon_n$, Twiss $\beta$-function at the waist $\beta^*$, and corresponding beam size $\sigma^*$ of the reconstructed transverse phase space distribution.}
    \label{tab:emi}
\end{table}

\section{\label{sec:dis}Discussion}
The reconstructed phase space represents the average distribution of many shots, since shot-to-shot fluctuations in the density cannot be characterized with multi-shot measurements like wire scans. Errors induced by total bunch charge fluctuations and position jitter of the electron beam could be corrected for by evaluating beam-synchronous BPM data. Since the BPMs in the Athos branch were still uncalibrated, their precision was insufficient to correct orbit jitter in our measurement. This issue is considered further in Appendix~\ref{app:err}. \\
The expected waist is located at the center of the chamber ($z=\SI{0}{cm}$), whereas the reconstructed waist is found \SI{6.2}{cm} downstream. In addition, the $\beta$-function at the waist ($\beta^*$) was measured to be around \SI{3.6}{cm} in both planes, which is in disagreement with the design optics ($\beta^*_x = \SI{1}{cm}$, $\beta^*_y = \SI{1.8}{cm}$). This indicates that the beam is mismatched at the chamber entrance and improving the matching of the electron beam to the focusing lattice could provide even smaller (sub-micrometer) beams in the ACHIP chamber. \\
The reconstructed normalized emittances are up to a factor of two larger than the normalized emittances measured after the second bunch compressor. This emittance increase can be attributed to various reasons. Within a distance of \SI{103}{m} the electron beam is accelerated from \SI{2.3}{GeV} (conventional emittance measurement) to around \SI{3.2}{GeV} and is directed to the Athos branch with a fast kicker and a series of bending magnets. Chromatic effects in the lattice, transverse offsets in the accelerating cavities or leaking dispersion from dispersive sections in the switch-yard can lead to a degradation of the emittance along the accelerator. These effects were not precisely characterized and corrected before the measurement, since the priority was to validate a new method for transverse phase space characterization of a strongly focused ultra-relativistic electron beam.\\
Another possible explanation for the discrepancy of the emittances: the conventional emittance measurement uses the horizontal and vertical beam profiles measured for different phase advances (quadrupole currents) with a scintillating screen (single-shot). 
A Gaussian fit to the beam profiles at each phase advance is used to estimate the emittance~\cite{prat2014symmetric}. In contrast, the tomographic wire scan technique presented here reconstructs the transverse phase space averaged over many shots. Afterwards, a Gaussian fit estimates the area of the distribution in the transverse phase space. Both large shot-to-shot jitter and non-Gaussian beams can give rise to differences between the results of the two techniques.\\
The wire scan acquisition time could be reduced by using fewer projection angles. This could be done, if less detailed information on the beam distribution is acceptable, e.g., if only projected beam sizes are of interest, two projection angles are sufficient. The optimal number of angles depends on the internal beam structure and the beam quantities of interest.
\subsection{Resolution Limit}
The ultimate resolution limit of the presented tomographic characterization of the transverse beam profile depends on the roughness of the wire profile. With the current fabrication process, this is on the order of \SI{100}{nm} estimated from electron microscope images of the free-standing gold wires. This is one to two orders of magnitude below the resolution of standard profile monitors for ultra-relativistic electron beams (YAG:Ce screens)~\cite{ischebeck2015transverse, maxson2017direct}. 
\subsection{Comparison to other Profile Monitors}
The scintillating screens (YAG:Ce) at SwissFEL achieve an optical resolution of \SI{8}{\micro m}, and the smallest measured beam sizes are \SI{15}{\micro m}~\cite{ischebeck2015transverse}. At the Pegasus Laboratory at UCLA beam sizes down to \SI{5}{\micro m} were measured with a \SI{20}{\micro m} thick YAG:Ce screen in combination with an in-vacuum microscope objective~\cite{maxson2017direct}. 
Optical transition radiation (OTR) based profile monitors are only limited by the optics and camera resolution~\cite{tenenbaum1999measurement}. At the Accelerator Test Facility 2 at KEK this technique was used to measure a beam size of \SI{750}{nm}~\cite{bolzon2015very}. However, OTR profile monitors are not suitable for compressed electron bunches (e.g., at FELs) due to the emission of coherent OTR~\cite{akre2008commissioning}. \\
At the SLAC Final Focus Test Beam experiment a laser-Compton monitor was used to characterize a \SI{70}{nm} wide beam along one dimension~\cite{balakin1995focusing}. The cost and complexity of this system, especially for multi-angle measurements, are its main draw-backs.\\
Concerning radiation hardness of the nano-fabricated wire scanner, tests with a single wire and a bunch charge of \SI{200}{pC} at a beam energy of \SI{300}{MeV} at SwissFEL did not show any sign of degradation after repeated measurements~\cite{orlandi2020nanofabricated}.

\section{\label{sec:con}Conclusion}
In summary, we have presented and validated a novel technique for the reconstruction of the transverse phase space of a strongly focused, ultra-relativistic electron beam. The method is based on a series of wire scans at different angles and positions along the waist. An iterative tomographic algorithm has been developed to reconstruct the transverse phase space.
The technique is validated with experimental data obtained in the ACHIP chamber at SwissFEL.
The method could be applied to other facilities and experiments, where focused high-brightness electron beams need to be characterized, for instance at plasma acceleration or DLA experiments for matching of an externally injected electron beam, emittance measurements at future compact low-emittance FELs~\cite{rosenzweig2020}, or for the characterization of the final-focus system at a high-energy collider test facility. For the latter application, the damage threshold of the free-standing nano-fabricated gold wires needs to be identified and radiation protection for the intense shower of scattered particles needs to be considered. Nevertheless, the focusing optics could be characterized with the presented method using a reduced bunch charge.

\begin{acknowledgments}
We would like to express our gratitude to the SwissFEL operations crew, the PSI expert groups, and the entire ACHIP collaboration for their support with these experiments.
We would like to thank Thomas Schietinger for careful proofreading of the manuscript.
This research is supported by the Gordon and Betty Moore Foundation through Grant GBMF4744 (ACHIP) to Stanford University.
\end{acknowledgments}
\appendix

\section{\label{app:err}Error Estimation}
\subsection{Position Errors}
The uncertainty of the position of the wire scanner with respect to the electron beam is affected by the readout precision of the hexapod ($<\SI{1}{nm}$), vibrational motion of the hexapod ($<\SI{10}{nm}$) and position jitter of the electron beam, which at SwissFEL is typically a few-percent of the beam size. 
The orbit of the electron beam is measured with BPMs along the accelerator. Unfortunately, the BPMs along the Athos branch of SwissFEL have not been calibrated (the measurement took place during the commissioning phase of Athos).
Nevertheless, we tried correcting the orbit shot-by-shot based on five BPMs and the magnetic lattice around the interaction point. However, it does not reduce the measured beam emittance, as their position reading is not precise enough to correct orbit jitter at the wire scanner location correctly.
Therefore, we do not include corrections to the wire positions based on BPMs. The reconstructed beam phase space represents the average distribution for many shots including orbit fluctuations. 
After the calibration of the BPMs in Athos we plan to characterize the effect of orbit jitter to wire scan measurements in detail. 
\subsection{Amplitude Errors}
Jitter to the BLM signal is introduced by read-out noise of the PMT ($<\SI{1}{\percent}$), charge fluctuations of the machine and halo-particles scattering at other elements of the accelerator. The charge measured by the BPMs fluctuated by \SI{1.3}{\percent} (rms) during the measurement. The signal-to-noise ratio (SNR) of the measurements varies from 25 to 45 depending on the respective projected beam size. We define the SNR as: $s_{\mathrm{max}}/\sigma_{\mathrm{noise}}$, where $s_{\mathrm{max}}$ is the maximum of the signal and $\sigma_{\mathrm{noise}}$ refers to the standard deviation of the background. 
\subsection{Uncertainty of the Reconstruction}
Due to the error sources mentioned above the measured projections are not fully compatible with each other, i.e., the reconstructed distribution cannot match to all measured data points.
The error of the reconstructed phase space density and the derived quantities is estimated by a procedure similar to the main reconstruction algorithm. The reconstructed distribution is now taken as input. Instead of averaging over all projections, the iteration is performed for each projection individually. Hence, a set of $n_z \times n_{\theta}$ distributions is generated, in which each distribution matches best to one measured projection. All derived quantities, such as the emittance or $\beta$-function, are computed for each distribution and the error is taken as the standard deviation of this set.\\

\begin{figure*}
\includegraphics[width=1.05\textwidth]{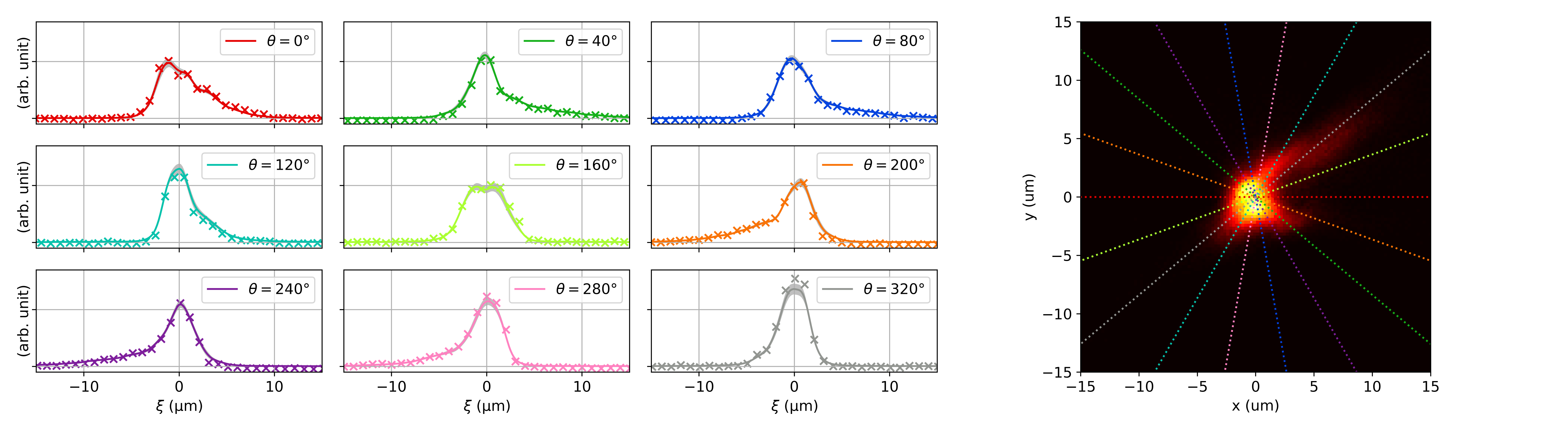}
\caption{\label{fig:10pCtomo} Tomographic reconstruction of a beam with non-Gaussian tails. The nine measured projections are indicated by crosses in the small nine sub-plots. The reconstruction result is shown in the larger sub-plot on the right ($x$,$y$ profile). The projections of the reconstruction are shown as solid lines in the corresponding sub-plots. The colors correspond to the different projection angles as indicated by dashed lines in the 2D profile plot on the right. The tomographic reconstruction is able to represent the core and tails of the beam.}
\end{figure*}
\begin{figure*}
\includegraphics[width=1.05\textwidth]{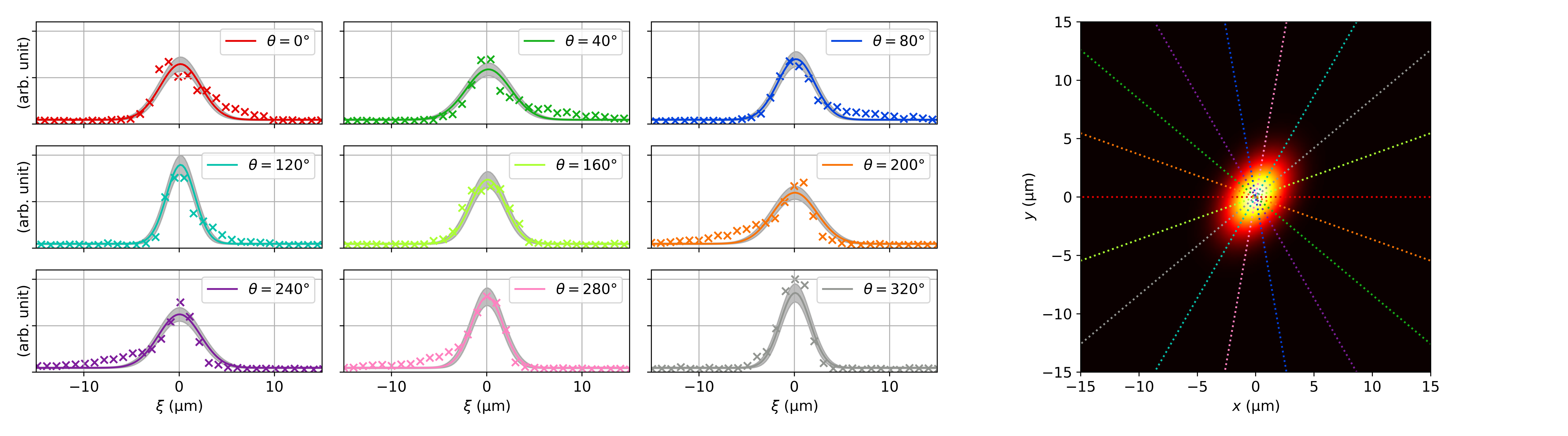}
\caption{\label{fig:10pCgauss} The result of a single Two-dimensional Gaussian fit to approximate nine measured projections. The measurement and the beam profile are shown analogously to the tomographic result shown in Fig.~\ref{fig:10pCtomo}. In contrast to the tomographic reconstruction, the Gaussian fit is not able to represent the tails correctly.}
\end{figure*}

\section{\label{app:term}Termination Criterion for Reconstruction Algorithm}
The algorithm to reconstruct the phase space from wire scan measurements iteratively approximates the distribution that fits best to all measurements (see Sec.~\ref{sec:alg}). The iteration is stopped when a criterion based on the relative change from the current to the previous iteration is reached. We define $p_k$ as the average probability for a particle to be added or removed to the ensemble in iteration $k$.
\begin{equation}
    p_k = \frac{1}{n_p n_{\theta} n_z} \sum_{i, \theta, z} |\Delta_{z, \theta}^i|
\end{equation}
The iteration terminates when the relative change of $p_k$ reaches a tolerance limit $\tau$:
\begin{equation}
    \frac{|p_k-p_{k-1}|}{|p_k|} < \tau
\end{equation}
For the case of the presented data set $\tau$ = 0.005 is found to provide stable convergence and a consistent solution. Around 110 iterations are required to reach the termination criterion.

\section{\label{app:nonGaussian} Reconstruction of non-Gaussian Beams}
Our particle based tomographic reconstruction algorithm does not assume any specific shape for the density profile. Therefore, asymmetric density variations, such as tails of a localized core can be reconstructed. To demonstrate this capability of our tomographic technique, we show here a measurement of a non-Gaussian beam shape and compare the result to a 2D Gaussian fit. This measurement was performed with different machine settings than the measurement presented in Sec.~\ref{sec:res}. The electron bunch carried a charge of around \SI{10}{pC}. The transverse beam profile was characterized with nine wire scans at different angles at one $z$ position. Therefore we can only reconstruct the two-dimensional ($x$,$y$) beam profile. The measurement and the tomographic reconstruction are shown in Fig.~\ref{fig:10pCtomo}. For comparison, we add the result of a single two-dimensional Gaussian fit to all nine measured projections (Fig.~\ref{fig:10pCgauss}). 
The core and tails observed in the measurement are well represented by the tomographic reconstruction, whereas the Gaussian fit overestimates the core region by trying to approximate the tails.

\bibliography{apssamp}

\end{document}